# Photon Position Operator and Localization of Photons inside a Waveguide

**Zhi-Yong Wang[1]\***, **Cai-Dong Xiong[1]**, **Ole Keller[2]**

[1]*Institute of Physical Electronics, UESTC, Chengdu, Sichuan 610054, P.R. China*
[2]*Institute of Physics, Aalborg University, Pontoppidanstrcede 103, DK-9220 AalborgØst, Denmark*

**Abstract**

In this article, we show that in the level of quantum mechanics, a photon position operator with commuting components can be obtained in a more natural way; in the level of quantum field theory, the photon position operator corresponds to the center of the photon number. It is most interesting for us to show that, a photon inside a waveguide can be localized in the same sense that a massive particle can be localized in free space, and just as that it is impossible to localize a massive particle with a greater precision than its Compton wavelength, one can also arrival at a similar conclusion: it is impossible to localize a photon inside a waveguide with a greater precision than its equivalent Compton wavelength, which owing to evanescent-wave or photon-tunneling phenomena.
.

\* Electronic address: zywang@uestc.edu.cn

# 1 Introduction

It has been proved that there is no self-adjoint position operator in the state space of the photon [1-2], which is taken to indicate that photons are not localizable at all. However, both possible theoretical objections and the experiment itself seem to reject such a radical interpretation of the mentioned results [3-14]. Consequently, in spite of the well-known conclusion that single photon cannot be localized, there have been many attempts to introduce the concept of photon wave function and reconstruct a first and second quantization description of the photon [3-9], and a photon position operator with commuting components has been constructed [10]. In fact, the conclusion of the nonlocalizability of the photon is disappointed from the perspective of quantum optics, and is hardly acceptable because it conflicts with our findings in laboratories [4]. For example, the optical resolution problem cannot be understood unless the spatial photon localization problem is addressed. What is more, some recent studies have shown that photons can be localized in space [11-13], where the Ref. [12] has shown that photons can be localized in space with an exponential falloff of the energy density and photon detection rate. Nowadays, a fresh interest in the photon localization problem has been awakened, meeting the needs of developments in near-field optics, cavity QED, and quantum computing.

In view of the mentioned above, the efforts of investigating the photon localization problem are valuable. In this article, we firstly construct photon position operators in the levels of quantum mechanics and quantum field theory, which is discussed in our framework of photon wave mechanics, and then investigate the localization of photons inside a waveguide (a rectangular waveguide without loss of generality), which is mainly based on the traditional framework of electromagnetic field theory.

In the following we apply the natural units of measurement ($\hbar = c = 1$) and the space-time metric tensor is taken as $g_{\mu\nu} = \text{diag}(1,-1,-1,-1)$. Repeated indices must be summed according to the Einstein rule. In this article we let $x^\mu = (t, -\boldsymbol{x})$, instead of $x^\mu = (t, \boldsymbol{x})$, denote the contravariant position 4-vector (and so on), then in our case $\hat{p}_\mu = i\dfrac{\partial}{\partial x^\mu} \equiv i\partial_\mu = i(\partial_t, -\nabla)$ denote 4-momentum operators ($\mu, \nu = 0, 1, 2, 3$), etc.

## 2  General points of view on the problem of position operator

In view of the fact that the quantum theory of the photon is directly a relativistic one, the discussion of the photon localization problem requires the investigation of defining position operators in relativistic quantum mechanics. The transition from nonrelativistic to relativistic quantum mechanics implies that the concepts of the nonrelativistic position operator have to be reinvestigated, this is because: 1) spatial and temporal coordinates have to be treated equally within relativistic quantum mechanics; 2) a relativistic particle cannot be localized more accurately than its Compton wavelength, otherwise the particle automatically has companions due to pair creation. There not all operators of the nonrelativistic theory can be transferred to the relativistic one-particle theory, and thus in relativistic quantum mechanics the consistency of the one-particle description is limited [15].

However, from nonrelativistic to relativistic theory there exists a unified point of view on localization problem. That is, even if expected from nonrelativistic quantum mechanics, the eigenfunctions of a position operator $\hat{\boldsymbol{x}}$ cannot be $\delta(\boldsymbol{x} - \boldsymbol{x}')$, which due to the fact that when a spatial interval becomes smaller and smaller, nonrelativistic quantum mechanics being tenable only at low energies becomes no longer valid. In more detail, as for a particle, the eigenfunctions of a position operator $\hat{\boldsymbol{x}}$ in position representation are not $\delta(\boldsymbol{x}-\boldsymbol{x}')$, but a kind of smeared-out delta function which is smeared out over a region of the order of the Compton wavelength of the particle. After all, the nonrelativistic theory is a low-energy approximation of the relativistic theory, and then the physical rules obeyed by the nonrelativistic theory belong to a subset of the physical rules obeyed by the relativistic theory. In a word, a position operator can be defined only in the limit of having no pair creation and destruction, or in the limit of measurement distance greater than the Compton wavelength. Correspondingly, one never achieves exact localization in nature.

A natural consider might be that, in the realm of relativistic quantum mechanics the concept of localizability of physical systems should be connected to the space-time notion, the one as a four-dimensional (4D) continuum. Correspondingly, one should view space-time as a whole and try to introduce a four-vector position operator, where time coordinate is treated as an operator on an equal footing with 3D spatial position operators. However, such consider raises several questions. For example, According to Pauli's argument [16], the existence of a self-adjoint time operator is incompatible with the semi-bounded character of the Hamiltonian spectrum. A way out of these dilemmas is based on the use of positive operator valued measures (POVMs) [17]. In fact, up to now, a variety of attempts have been made to resolve the position-operator problem within the framework of relativistic quantum mechanics and quantum field theory, where the most important attempts are those via introducing a four-vector position operators [18-21] and using the formalism of POVMs [21-22, 14].

Here our main purpose is to investigate the 3D spatial position operator and localization of the photon (as the massless particle).

## 3  Photon position operator in quantum-mechanical level

Let $k^\mu = (\omega, -\boldsymbol{k})$ stand for the 4-momentum of a single Fourier mode of the photon field ($\hbar = c = 1$), where $\omega = k^0$ is also the angular frequency and $\boldsymbol{k}$ also the wave number vector. In Ref. [23], Maxwell's theory in vacuum medium is restated in a form closely parallels to Dirac's theory of the electron, where the free photon field is described as a 6×1 spinor $\psi(x)$ that transforms according to the $(1,0) \oplus (0,1)$ representation of the Lorentz group (the symbol $\oplus$ represents the direct sum), where the corresponding equation (say, the *Dirac-like equation*) has a Lorentz-covariant form. In such a way, photon wave mechanics is reinvestigated in a more rigorous and unified way, where the Dirac-like equation provide an unified and complete description for all kinds of electromagnetic fields outside a source, including the time-varying and static fields, the transverse and longitudinal fields, and those generated by an electrical or magnetic source, etc. The 6×1 spinor $\psi(x)$ can be written as, generally

$$\psi(x) = \sum_{k,\lambda} \sqrt{\frac{\omega}{2V}} [a(k,\lambda) f(k,\lambda) \exp(-\mathrm{i} k \cdot x) + b^+(k,\lambda) g(k,\lambda) \exp(\mathrm{i} k \cdot x)] \quad (1)$$

where $\lambda = \pm 1, 0$, the expansion coefficients $a(k,\lambda)$ and $b^+(k,\lambda)$ (the Hermitian conjugate of $b(k,\lambda)$) are the annihilation and creation operators, and

$$f(k,\lambda) = \frac{1}{\sqrt{1+\lambda^2}} \begin{pmatrix} \varepsilon(\boldsymbol{k},\lambda) \\ \lambda \varepsilon(\boldsymbol{k},\lambda) \end{pmatrix}, \quad g(k,\lambda) = \frac{1}{\sqrt{1+\lambda^2}} \begin{pmatrix} \lambda \varepsilon(\boldsymbol{k},\lambda) \\ \varepsilon(\boldsymbol{k},\lambda) \end{pmatrix} \quad (2)$$

where three unit polarization vectors $\varepsilon(\boldsymbol{k},\lambda)$ (in the form of 3×1 column matrices, $\lambda = \pm 1, 0$) satisfy the orthonormal and completeness relations

$$\begin{cases} \varepsilon^+(\boldsymbol{k},\lambda)\varepsilon(\boldsymbol{k},\lambda') = \delta_{\lambda\lambda'} \\ \sum_\lambda \varepsilon(\boldsymbol{k},\lambda)\varepsilon^+(\boldsymbol{k},\lambda) = I_{3\times 3} \end{cases} \quad (3)$$

According to Ref. [23], by $(-i\partial_\mu)^a \exp(\pm ik\cdot x) = (\pm k_\mu)^a \exp(\pm ik\cdot x)$ one defines the arbitrary-order derivative, and the scalar product between two spinors is assumed to be of the form

$$\langle \psi_1 | \psi_2 \rangle = \int d^3 x [(-i\partial_t)^{-1} \psi_1^+(x)] \psi_2(x) \quad (4)$$

with the assumption that the periodic boundary conditions for $\psi(x)$ enclosed in a box taken to be a cube of side $\sqrt[3]{V}$, in momentum space, the scalar product corresponding to Eq. (4) can be expressed as

$$\langle \varphi_1 | \varphi_2 \rangle = \sum_k \frac{1}{\omega} \varphi_1^+(k) \varphi_2(k) \quad (5)$$

where $\varphi_1(k)$ and $\varphi_2(k)$ has the dimension of $[1/\text{length}]^{1/2}$ such that the scalar product is dimensionless. As we known, an operator has different expression in different representation-space. As an example, consider that the momentum-space unit spinors $f(k,\lambda)$ and $g(k,\lambda)$ both are constructed by $\varepsilon(\boldsymbol{k},\lambda)$, using Eq. (5) a momentum-space wavefunction can be chosen as

$$\varphi_{\boldsymbol{x}_0,\lambda}(k) = C_\lambda \sqrt{\omega}\, \varepsilon(\boldsymbol{k},\lambda) \exp(-i\boldsymbol{x}_0 \cdot \boldsymbol{k}) \quad (6)$$

where $C_\lambda$ is a dimensionless constant. Consider that in this article we set $x^\mu = (t, -\mathbf{x})$ and $\hat{p}_\mu = i\dfrac{\partial}{\partial x^\mu} \equiv i\partial_\mu = i(\partial_t, -\nabla)$, the naive position operator is $i\nabla_k \equiv i(\dfrac{\partial}{\partial k^1}, \dfrac{\partial}{\partial k^2}, \dfrac{\partial}{\partial k^3})$, using Eq. (3), one has

$$\nabla_k \varepsilon(\mathbf{k},\lambda) = \sum_{\lambda'}[\nabla_k \varepsilon(\mathbf{k},\lambda')]\varepsilon^+(\mathbf{k},\lambda')\varepsilon(\mathbf{k},\lambda) \quad (7)$$

Let the naive position operator act on $\varphi_{\bar{x}_0,\lambda}(k)$, using $\omega = \sqrt{k_1^2 + k_2^2 + k_3^2}$ one has

$$\hat{x}\varphi_{x_0,\lambda}(k) = x_0 \varphi_{x_0,\lambda}(k) \quad (8)$$

Let $I_{3\times 3}$ denote the 3×3 unit matrix, one has

$$\hat{x} = i(\nabla_k - \frac{\mathbf{k}}{2\omega^2})I_{3\times 3} - i\sum_\lambda [\nabla_k \varepsilon(\mathbf{k},\lambda)]\varepsilon^+(\mathbf{k},\lambda) \quad (9)$$

Then $\varphi_{\bar{x}_0,\lambda}(k)$ is an eigenfunction of $\hat{x}$ with the eigenvalue $x_0$. In fact, one can regard $\hat{x}$ as the photon position operator with commuting components, which is the same as that given in Ref. [10]. However, it is worth stressing that:

1) As compared with Ref. [10], the result obtained here is more natural (because the $\lambda = 0$ solutions are not discarded in our case), while in Ref. [10], the reason that the three-vector wave functions in momentum-space also contain the longitudinal component seems to be insufficient.

2) As compared with Ref. [14], here the presence of longitudinal field components does not cause any trouble, while in Ref. [14], in order to obtain a photon position operator with commuting components, the author has constructed an enlarged Hilbert space that also contains the longitudinal field components.

**4　Photon position operator in the level of quantum field theory**

In processes that do not involve particle-pair creation or annihilation, a position operator can be defined within the second-quantization formalism. In the present case, the photon position operator can be defined as, in the level of quantum field theory

$$X = \int [(-i\partial_t)^{-1}\psi^+(x)] x \psi(x) \mathrm{d}^3 x \qquad (10)$$

In the following we shall make no distinction between the photons come from electric multipole moments and those from magnetic multipole moments, and thus let $b(k,\lambda) = a(k,\lambda)$ in the final results. Applying Eq. (1), one has

$$\begin{aligned}x\psi(x) &= \sum_{k,\lambda}\sqrt{\frac{\omega}{2V}}[a(k,\lambda)f(k,\lambda)x\mathrm{e}^{-ik\cdot x} + a^+(k,\lambda)g(k,\lambda)x\mathrm{e}^{ik\cdot x}] \\ &= \sum_{k,\lambda}\sqrt{\frac{\omega}{2V}}[a(k,\lambda)f(k,\lambda)(i\nabla_k)\mathrm{e}^{-ik\cdot x} - a^+(k,\lambda)g(k,\lambda)(i\nabla_k)\mathrm{e}^{ik\cdot x}]\end{aligned} \qquad (11)$$

Consider that $\int \mathrm{d}^3 x \exp[i(\bm{k}-\bm{k}')\cdot x] = V\delta_{kk'}$, $[a(k,\lambda), a^+(k',\lambda')] = \delta_{kk'}\delta_{\lambda\lambda'}$, using Eqs. (2)-(3), one can obtain

$$X = \sum_{k,\lambda} a^+(k,\lambda) a(k,\lambda)(i\nabla_k) \qquad (12)$$

Therefore we can interpret the position operator $X$ as a center of the photon number. It is interesting to note that Eq. (12) can be also interpreted as the position-quantization of the electromagnetic field, which is due to the fact that the electromagnetic field consists of photons. In fact, starting from Dirac electron theory, in a similar way one can obtain the counterpart of Eq. (12).

If $a(k,\lambda)$ and $b(k,\lambda)$ are independent of $k$, consider that

$$\begin{aligned}\sum_{k,\lambda}\sqrt{\omega}b^+(k,\lambda)g(k,\lambda)\mathrm{e}^{i(\omega t - \bm{k}\cdot x)} &= \sum_{k,\lambda}\sqrt{\omega}b^+(k_0,-\bm{k},\lambda)g(k_0,-\bm{k},\lambda)\mathrm{e}^{i(\omega t + \bm{k}\cdot x)} \\ \sqrt{\omega}f(k,\lambda)x\mathrm{e}^{-i(\omega t - \bm{k}\cdot x)} &= \hat{X}_+ \sqrt{\omega}f(k,\lambda)\mathrm{e}^{-i(\omega t - \bm{k}\cdot x)} \\ \sqrt{\omega}g(k_0,-\bm{k},\lambda)x\mathrm{e}^{i(\omega t + \bm{k}\cdot x)} &= \hat{X}_- \sqrt{\omega}f(k_0,-\bm{k},\lambda)\mathrm{e}^{i(\omega t + \bm{k}\cdot x)}\end{aligned} \qquad (13)$$

where

$$\hat{X}_+ = \mathrm{i}(\nabla_k - \frac{k}{2\omega^2})I_{6\times 6} - \mathrm{i}\sum_\lambda [\nabla_k f(k,\lambda)]f^+(k,\lambda)$$
$$\hat{X}_- = \mathrm{i}(\nabla_k - \frac{k}{2\omega^2})I_{6\times 6} - \mathrm{i}\sum_\lambda [\nabla_k g(k_0,-k,\lambda)]g^+(k_0,-k,\lambda)$$
(14)

In the present case, Eq. (10) becomes as

$$X = \sum_k \frac{1}{2\omega}[\varphi_+^+(k)\hat{X}_+\varphi_+(k) + \varphi_-^+(k)\hat{X}_-\varphi_-(k)] \quad (15)$$

where

$$\varphi_+(k) = \sum_\lambda \sqrt{\omega}a(k,\lambda)f(k,\lambda)$$
$$\varphi_-(k) = \sum_\lambda \sqrt{\omega}b^+(k_0,-k,\lambda)g(k_0,-k,\lambda)$$
(16)

can be regarded as the momentum-space field operators with the positive- and negative-frequencies, respectively. In our case, the momentum-space scalar product is defined by Eq. (5), while Eq. (15) represents the momentum-space average of the photon position operator.

Therefore, in the level of quantum field theory, the photon position operator $X$ is directly related to the naive position operator $\mathrm{i}\nabla_k$ via Eq. (12) and can be interpreted as a center of the photon number. If the numbers of the photons with different momentums are the same, the photon position operator (see Eqs. (14)-(15)) corresponds to the momentum-space average of the operators that similar to the Hawton's position operator presented in Ref. [10].

## 5 Localization of photons inside a waveguide

People have shown that a single photon cannot be localized in the same sense that a massive particle in nonrelativistic quantum mechanics can be localized [24]. However, not only the massless particle such as the photon, but also the massive particle itself, cannot be localized in the traditional sense that a massive particle in nonrelativistic quantum mechanics can be localized. This is due to the fact that, when a particle is confined to a small region, nonrelativistic quantum mechanics may become no longer valid. Therefore, the concept of localizability in nonrelativistic quantum mechanics is just an ideal one. On the other hand, in the level of relativistic quantum mechanics, there has a common property for the localizability of all massive particles: it is impossible to localize a particle with a greater precision than its Compton wavelength, which owing to many-particle phenomena. Therefore, though there has no nonrelativistic quantum mechanics of the photon, it presents no problem; one can study the localizability of the photon only in the sense that a massive particle in relativistic quantum mechanics can be localized.

In practice, an interaction is necessary to locate a particle. On the other hand, in all the electromagnetism techniques, an interaction between photons and the related apparatuses is inevitable. These interactions include that of confining photons to a spatial region, where the dimension of the region can be 1, 2, or 3. As an example, we shall show that the photons inside a waveguide can be localized in the same sense that a massive particle can be localized in free space.

In a Cartesian coordinate system spanned by an orthonormal basis $\{e_1, e_2, e_3\}$ with $e_3 = e_1 \times e_2$, we assume that a hollow metallic waveguide is placed along the direction of $e_3$, and the waveguide is a straight rectangular pipe with the transversal dimensions being $b_1$ and $b_2$, respectively, and let $b_1 > b_2$ without loss of generality. The conductivity of the metal pipe is assumed to be infinite. It is also assumed that the waveguide is infinitely long and the source of the electromagnetic field is localized at infinity. The pipe is considered as a boundary which confines the waves to the enclosed space.

When the electromagnetic waves propagate along the direction of the waveguide, they are reflected back and forth by perfectly conducting walls in the empty inner space of the pipe. For simplicity, we assume that the photons inside the waveguide correspond to a given eigenmode of the waveguide, and have 4-momentum $k_\mu = (\omega, \boldsymbol{k})$ in the Cartesian coordinate system $\{\boldsymbol{e}_1, \boldsymbol{e}_2, \boldsymbol{e}_3\}$, where the wave-number vector is $\boldsymbol{k} = \sum_{i=1}^{3} \boldsymbol{e}_i k_i = (k_1, k_2, k_3)$ and the frequency $\omega = |\boldsymbol{k}| = \sqrt{k_1^2 + k_2^2 + k_3^2}$. Consider that the waveguide is placed along the direction of $\boldsymbol{e}_3$, the allowed values for the wave numbers $k_1$ and $k_2$ are fixed by $b_1$ and $b_2$, respectively, that is, $k_1 = r\pi/b_1$ and $k_2 = s\pi/b_2$, where $r = 1, 2, 3...$, $s = 0, 1, 2...$. Then the cutoff frequency of the waveguide is $\omega_c = \sqrt{k_1^2 + k_2^2}$ ($\hbar = c = 1$), which is determined by the transversal dimensions of the waveguide.

We define the *apparent mass* of photons inside the waveguide as

$$m = \hbar \omega_c / c^2 = \omega_c = \sqrt{k_1^2 + k_2^2} \quad (17)$$

Then the photon energy $E = \omega$ satisfies

$$E^2 = k_3^2 + m^2 \quad (18)$$

Now we choose another Cartesian coordinate system formed by an orthonormal basis $\{\boldsymbol{a}_1, \boldsymbol{a}_2, \boldsymbol{a}_3\}$ with $\boldsymbol{a}_3 = \boldsymbol{a}_1 \times \boldsymbol{a}_2$, such that in the new coordinate system, one has

$$\boldsymbol{e}_3 k_3 = \sum_{i=1}^{3} \boldsymbol{a}_i p_i = \boldsymbol{p} \quad (19)$$

In the following we let $k_3 \geq 0$ without loss of generality. If $k_3 = |p| > 0$, in the coordinate system $\{a_1, a_2, a_3\}$, the unit vectors $e_i$ ($i = 1, 2, 3$) can be expressed as

$$\begin{cases} e_1 = e(p,1) = (\dfrac{p_1^2 p_3 + p_2^2 |p|}{|p|(p_1^2 + p_2^2)}, \dfrac{p_1 p_2 p_3 - p_1 p_2 |p|}{|p|(p_1^2 + p_2^2)}, -\dfrac{p_1}{|p|}) \\ e_2 = e(p,2) = (\dfrac{p_1 p_2 p_3 - p_1 p_2 |p|}{|p|(p_1^2 + p_2^2)}, \dfrac{p_2^2 p_3 + p_1^2 |p|}{|p|(p_1^2 + p_2^2)}, -\dfrac{p_2}{|p|}) \\ e_3 = e(p,3) = e(p,1) \times e(p,2) = \dfrac{p}{|p|} = \dfrac{1}{|p|}(p_1, p_2, p_3) \end{cases} \quad (20)$$

Obviously, as $p_3 \to |p| = k_3$ such that $p \to (0, 0, p_3)$, one has $(e_i)_j = \delta_{ij}$, i.e., the coordinate system $\{a_1, a_2, a_3\}$ coincides with $\{e_1, e_2, e_3\}$ by $a_i = e_i$ ($i, j = 1, 2, 3$). Note that, in the Cartesian coordinate system $\{a_1, a_2, a_3\}$, the frequency $\omega$ is invariant (i.e., in a 3D spatial rotation that results in a transformation $\{e_1, e_2, e_3\} \to \{a_1, a_2, a_3\}$, the time component $\omega$ of the 4-momentum $k_\mu = (\omega, k)$ would keep invariant). What is more, in the coordinate system $\{a_1, a_2, a_3\}$, the momentum 3-vector of the photons inside the waveguide can be written as

$$k = k_T + k_L \quad (21)$$

where

$$k_T \equiv e(p,1) k_1 + e(p,2) k_2, \quad k_L = p = \sum_{i=1}^{3} a_i p_i \quad (22)$$

stand for the transverse and longitudinal components of $k$, respectively. In the coordinate system $\{a_1, a_2, a_3\}$, Eq. (18) can be expressed as the form of the usual relativistic energy-momentum relation (or, the relativistic dispersion relation):

$$E^2 = p^2 + m^2 \quad (23)$$

this is closely analogous to that of a material (massive) particle. Then, the cutoff frequency $\omega_c$, as the vertical-motion energy of the photons inside the waveguide, plays the role of rest mass. Moreover, in the coordinate system $\{a_1, a_2, a_3\}$, the 3D vector $p = \sum_{i=1}^{3} a_i p_i$ represents the momentum of the photons along the direction of the waveguide, while the 3D vector $k_T$ is perpendicular to $p$ and then represents transverse momentum, it contributes the apparent mass $m$ to the photons in such a way:

$$m = \omega_c = |k_T| \qquad (17)'$$

That is, the constraint on the motion of the photons in the transverse direction of the waveguide gives rise to an equivalent mass to the photons.

Furthermore, the group velocity (denoted by $v_g$) and phase velocity (denoted by $v_p$) of the photons inside the waveguide are, respectively ($\hbar = c = 1$)

$$\begin{aligned} v_g &= e(p,3)\sqrt{1-(\omega_c/\omega)^2} = e(p,3)\frac{|p|}{E} = \frac{p}{E} \\ v_p &= \frac{e(p,3)}{\sqrt{1-(\omega_c/\omega)^2}} = e(p,3)\frac{E}{|p|} \end{aligned} \qquad (24)$$

Further, one can obtain the de Broglie's relations ($v_g = |v_g|$, $v_p = |v_p|$)

$$\begin{aligned} v_g v_p &= c^2 = 1 \\ E^2 &= p^2 c^2 + m^2 c^4 = p^2 + m^2 \\ E &= \hbar\omega = \omega \\ p &= \hbar k_g = k_g \end{aligned} \qquad (25)$$

where $k_g = e(p,3) 2\pi/\lambda_g$ is the guide-wavenumber vector, $\lambda_g$ is the actual guide wavelength inside the waveguide, which is defined as the distance between two equal phase planes along the waveguide and plays the role of the de Broglie's wavelength. In terms of the free-space wavelength $\lambda = 2\pi/|\mathbf{k}|$, one has $\lambda_g = \lambda/\sqrt{1-(\omega_c/\omega)^2}$. Form Eqs. (24)-(25) one can also obtain the relativistic energy

$$E = \frac{mc^2}{\sqrt{1-(v_g^2/c^2)}} = \frac{m}{\sqrt{1-v_g^2}} \qquad (26)$$

In fact, inside the waveguide, the group velocity $\mathbf{v}_g$ of the photons can be viewed as a relative velocity between an observer and the photons with the apparent mass $m = \omega_c$.

From Eqs. (23)-(26), one can show that there exists a closely analogy between the behaviors of the de Broglie matter waves and those of the electromagnetic waves inside a waveguide. In fact, the similarity between the dispersion relation of matter waves for material (massive) particles and that of electromagnetic waves inside the waveguide has been mentioned before [25-26].

To discuss these rigorously and generally, we are suggested by the concept of the apparent mass and Eq. (23), and define the apparent 4-momentum of the photons inside the waveguide as $k_{L\mu} \equiv (E, \mathbf{p})$. Furthermore, consider that the apparent mass is always positive, that is, $m = |\mathbf{k}_T| = \sqrt{(r\pi/b_1)^2 + (s\pi/b_2)^2} > 0$, one can define a space-like 4-vector $\eta_\mu \equiv (0, \mathbf{k}_T/|\mathbf{k}_T|)$ that obviously satisfies $\eta_\mu \eta^\mu = -1$. By $\eta_\mu$ we introduce a space-like 4-momentum $k_{T\mu} \equiv (0, \mathbf{k}_T) = m\eta_\mu$, which describes the motion along the direction perpendicular to the apparent four-momentum $k_{L\mu}$. In such a way, we obtain an orthogonal decomposition for the 4-momentum of the photons inside the waveguide, namely, one has

$$k_\mu = (\omega, \boldsymbol{k}) = k_{T\mu} + k_{L\mu}$$
$$k_{T\mu} \equiv (0, \boldsymbol{k}_T) = m\eta_\mu, \quad k_{L\mu} \equiv (E, \boldsymbol{p}) \quad (27)$$
$$k_{L\mu}k_T^\mu = 0, \quad \eta_\mu \eta^\mu = -1$$

where note that the orthogonal relation $k_{L\mu}k_T^\mu = 0$ is Lorentz invariant. In fact, all light-like 4-vectors can be orthogonally decomposed as the sum of a space-like 4-vector and a time-like 4-vector. As for the photons inside the waveguide, the time-like 4-momentum $k_{L\mu}$ representing their apparent 4-momentum, can be called the active part of $k_\mu$, or the kinematics part of $k_\mu$; while the space-like 4-momentum $k_T^\mu$ contributing the apparent mass term, can be called the frozen part of $k_\mu$, or the stationary-wave part of $k_\mu$. In a word, the constraint on the motion of the photons in the transverse direction of the waveguide gives rise to an apparent mass to the photons, or equivalently, the apparent mass term arises by freezing out the degrees of freedom corresponding to the transverse 4-momentum $k_T^\mu$.

Now, from the point of view of 4D space-time, we shall give a description of the electromagnetic waveguide theory in a Lorentz covariant formulation. As discussed before, in the Cartesian coordinate system $\{\boldsymbol{a}_1, \boldsymbol{a}_2, \boldsymbol{a}_3\}$, the waveguide is placed along the direction of the vector $\boldsymbol{p} = \sum_{i=1}^{3} \boldsymbol{a}_i p_i$. Consider that the divergence of electric field intensities must be zero in the free space inside the waveguide (since there are no charges there), in spite of the boundary conditions for the waveguide, the photons in the free space inside the waveguide should obey the Maxwell equations, or, the Dirac-like equation [23]

$$i\beta^\mu \partial'_\mu \psi'(x) = 0 \quad (28)$$

According to Ref. [23], in terms of $\boldsymbol{\tau} = (\tau_1, \tau_2, \tau_3)$, where $(\tau_l)_{mn} = -i\varepsilon_{lmn}$ ($\varepsilon_{lmn}$ denote the full antisymmetric tensor with $\varepsilon_{123} = 1$), the matrices $\beta^\mu = (\beta^0, -\boldsymbol{\beta})$ can be written as

$$\beta_0 = \begin{pmatrix} I_{3\times 3} & 0 \\ 0 & -I_{3\times 3} \end{pmatrix}, \quad \boldsymbol{\beta} = (\beta_1, \beta_2, \beta_3) = \begin{pmatrix} 0 & \boldsymbol{\tau} \\ -\boldsymbol{\tau} & 0 \end{pmatrix} \quad (29)$$

Corresponding to Eq. (27), i.e., the orthogonal decomposition of the 4-momentum of the photons inside the waveguide, one can make an orthogonal decomposition for the 4-momentum operator $i\partial'_\mu$ presented in Eq. (28)

$$i\partial'_\mu = i\partial_\mu + m\eta_\mu = i(\partial_\mu - im\eta_\mu) \quad (30)$$

such that Eq. (28) becomes as

$$i\beta^\mu(\partial_\mu - im\eta_\mu)\psi(x) = 0 \quad (31)$$

For simplicity, let $\psi'(x) = \varphi(k)\exp(-ik_\mu x^\mu)$ and hence $\psi'(x) = \psi(x)\exp(-ik_{T\mu}x^\mu)$, i.e., $\psi(x) = \varphi(k)\exp(-ik_{L\mu}x^\mu)$. In the present case, one has

$$\begin{aligned} i\partial'_\mu \psi'(x) &= k_\mu \psi'(x) \\ i\partial_\mu \psi(x) &= k_{L\mu} \psi(x) \\ \eta^\mu \partial_\mu \psi(x) &= \partial_\mu \eta^\mu \psi(x) = 0 \end{aligned} \quad (32)$$

Correspondingly, by replacing $\nabla'$ with $(\nabla - im\boldsymbol{\eta})$ one can obtain a new expression of the original transversality conditions. Using Eqs. (27), (31) and (32), and denote $\hat{p}_\mu = i\partial_\mu$, one can obtain

$$(\hat{p}_\mu \hat{p}^\mu - m^2)\psi(x) = 0 \quad (33)$$

or, in the momentum space

$$(k_{L\mu} k_L^\mu - m^2)\varphi(k) = (E^2 - \boldsymbol{p}^2 - m^2)\varphi(k) = 0 \quad (34)$$

which is agreement with Eq. (23). In fact, using Eq. (27), one can directly obtain

$$k_{\text{L}\mu}k_{\text{L}}^{\mu} - m^2 = k_{\text{L}\mu}k_{\text{L}}^{\mu} + k_{\text{T}\mu}k_{\text{T}}^{\mu} = k_{\mu}k^{\mu} = 0 \quad (35)$$

In a word, the photons in the free space inside the waveguide satisfy the new Dirac-like equation (31) and the Klein-Gordon equation (33), where the cutoff frequency $m = \omega_c$ plays the role of rest mass. Therefore, inside the waveguide, the electromagnetic waves for photons become a concept analogous to that of the matter waves for material (massive) particles, which can be shown further by Eqs. (23)-(26). This implies that the photons inside the waveguide may be localized in the usual sense that a massive particle can be localized in free space.

In fact, there has another way of looking at the discussions above. In detail, we shall show that, what we discuss above provides a realistic model for the theory given by Ref. [20]. In Ref. [20] the authors have introduced quantum observables describing the space-time position of the physical event defined by the intersection of two light pulses, where the conformal-transformation generators (related to conformal symmetry of massless quantum fields) is used first to build the definition of there space-time observables, by which the connection between the definition of an observable space-time position and the massive character of the state has been obtained a simple physical interpretation.

As we known, on the one hand, the TEM mode (both the electric and magnetic fields perpendicular to the direction of propagation) cannot propagate in a single conductor transmission line, only those modes in the form of transverse electric (TE) and transverse magnetic (TM) modes can propagate in the waveguide. On the other hand, however, there has another useful way of looking at the propagation [26-27], that is, the waveguide field can be viewed as the superposition of two sets of plane waves (i.e., the TEM waves) being continually reflected back and forth between perfectly conducting walls and zigzagging down the waveguide (see **Figure 1**). The two sets of plane waves have the same amplitudes and frequencies, but reverse phases. Namely, for a given wall of the waveguide, if one set of plane waves correspond to the incident waves, then another to the reflected wave (a reflection means reversal of phase). Let $k_{1\mu}$ and $k_{2\mu}$ denote the 4-momemnta of the two sets of plane waves, respectively, one has, obviously

$$k_{1\mu}k_1^\mu = k_{2\mu}k_2^\mu = 0$$
$$(k_{1\mu} + k_{2\mu})(k_1^\mu + k_2^\mu) = 4(E^2 - \boldsymbol{p}^2) = 4m^2 \qquad (36)$$

Where the factor of 4 presents no problem, because once the waveguide field is expressed as the superposition of two sets of plane waves, its amplitude is halved. Equation (36) implies that, with the aid of the Poincaré algebra one can construct the position operator of the photons inside the waveguide, in the same sense that the position operator of a massive particle can be constructed. The detailed discussions are similar to those presented in Ref. [20]. However, in contrast with Ref. [20], here one can show that, not only the presence of the two photons propagating in different directions, but also the superposition of two different wave-vector states of a single photon, allow one to introduce the definition of a localized event. The second case is due to the presence of the apparent mass of the photons inside the waveguide.

As mentioned before, there has a common property for the localizability of all massive particles: it is impossible to localize a particle with a greater precision than its Compton wavelength, which owing to many-particle phenomena. Likewise, in terms of the concept of the apparent mass $m = \omega_c$, one can define the equivalent Compton wavelength of the photons inside the waveguide

$$\lambda_{com} = \frac{\hbar}{\omega_c c} = \frac{\hbar}{mc} = \frac{1}{m} \qquad (37)$$

Similarly, as for the localizability of the photons inside the waveguide, one can arrival at such a conclusion: it is impossible to localize a photon inside a waveguide with a greater precision than its equivalent Compton wavelength, which owing to evanescent-wave or photon-tunneling phenomena. For example, assume that a photon inside a waveguide lies in a given mode of the waveguide, if one tries to reduce the size of the cross section of the waveguide and obtain a new waveguide, so as to localize the photon within the cross section with a greater precision than its equivalent Compton wavelength, then there always exists an inertial reference frame in which the photon's frequency is smaller than the cutoff frequency of the new waveguide, this shows that the photon-tunneling phenomenon would take place. In fact, here we have given another way of looking at the relation between optical tunneling and photon localizability, which has been discussed in Ref. [28]

## 6  Conclusions

In the level of quantum mechanics, a photon position operator with commuting components can be obtained in a more natural way as compared with Refs. [10] and [14]; in the level of quantum field theory, we find that the photon position operator (given by Eq. (12)) corresponds to the average of the naive position operator with respect to the particle numbers and can be interpreted as a center of the photon number. It is interesting to note that Eq. (12) can be also interpreted as the position-quantization of the electromagnetic field, which is due to the fact that the electromagnetic field consists of photons. In fact, starting from Dirac electron theory, in a similar way one can obtain the counterpart of Eq. (12). It is most interesting for us to show that, a photon inside a waveguide can be localized in the same sense that a massive particle can be localized in free space, this actually provides a realistic model for the theory given by Ref. [20]. Just as that it is impossible to localize a massive particle with a greater precision than its Compton wavelength (due to many-particle phenomenon), one can also arrival at a similar conclusion: it is impossible to localize a photon inside a waveguide with a greater precision than its equivalent Compton wavelength, which owing to evanescent-wave or photon-tunneling phenomena.

**Acknowledgments**

The first author (Z. Y. Wang) would like to thank Dr Lu Changhai for helpful discussions. The authors wish to acknowledge financial support from China National Natural Science Foundation and the Excellent Young Teachers Program of MOE of China (No: 69971008).

**Figure Caption**

**Figure 1:** Wave propagation in a waveguide. The waveguide field can be viewed as the superposition of two trains of plane waves, where $\lambda_g$ is the guide wavelength, $\lambda$ the free-space wavelength, $\bar{v}_g$ the group velocity. Let $k_{1\mu}$ and $k_{2\mu}$ denote the 4-momemnta of the two sets of plane waves, they satisfy equation (36).

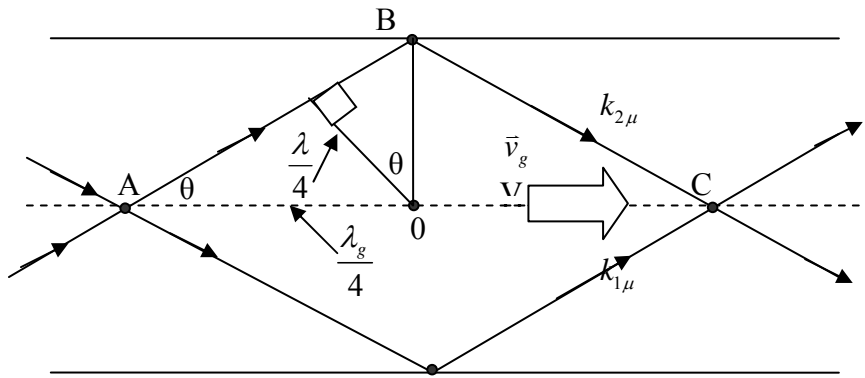

**Figure 1**

Wang *et al*